\documentclass[aps,preprint]{revtex4-2}

\usepackage{amssymb}
\usepackage{amsmath}
\usepackage{epsfig}
\usepackage{epstopdf}
\usepackage{bm}
\usepackage{graphicx,epsfig}
\usepackage{mathrsfs}
\usepackage{dcolumn}
\usepackage{color}
\usepackage{natbib}
\usepackage{CJK}
\usepackage{bm}
\allowdisplaybreaks[2]

\begin{document}

\title{Extreme narrow band in Moir\'e  Photonic time crystal}
\author{Zhaohui Dong$^{1}$, Xianfeng Chen$^{1,2}$, and Luqi Yuan$^{1,*}$}
\affiliation{$^1$State Key Laboratory of Advanced Optical Communication Systems and Networks, School of Physics and Astronomy, Shanghai Jiao Tong University, Shanghai 200240, China \\
$^2$Collaborative Innovation Center of Light Manipulation and Applications, Shandong Normal University, Jinan 250358, China \\
$^{\ast} $yuanluqi@sjtu.edu.cn}

\begin{abstract}
    The Moir\'e superlattice has attracted growing interest in the electromagnetic and optical communities. Here, we extend this concept to time-varying photonic systems by superposing two binary modulations on the refractive index with different modulation periods, i.e., the Moir\'e photonic time crystal (PTC). Such a Moir\'e PTC leads to extreme narrow bands in momentum space which supports temporal localized modes, exhibiting periodically self-reconstructing pulse in time domain. We investigate how the modulation parameters change the bandstructure of the Moir\'e PTC and the temporal localization behavior. Moreover, we explore mode-locking mechanism in frequency space in the Moir\'e PTC, which points towards potential applications in mode-locked lasers with tunable time width of the emitted pulses. Our work therefore extends the study of PTC to complex modulation patterns, and unveils new possibility in wave manipulation with time-varying systems.

\end{abstract}

\maketitle
The concept of Moir\'e superlattice provides a new degree of freedom in the design of photonic crystals \cite{du2023moire, oudich2024engineered, du2024nonlinear}, which brings intriguing opportunities in wave manipulation, such as localization of light field \cite{sunku2018photonic, wang2020localization, tang2021modeling, yi2022strong, ma2023twisted}, moiré nanolaser arrays \cite{mao2021magic, raun2023gan, luan2023reconfigurable}, optical solitons \cite{fu2020optical, kartashov2021multifrequency}, and moiré bound states in the continuum \cite{huang2022moire, qin2024optical}. 
Recently, there have been growing interest in the temporal counterpart of photonic crystals, i.e., PTCs \cite{lustig2023photonic, saha2023photonic, boltasseva2024photonic, asgari2024photonic}. 
Such systems hold spatial homogeneity while being periodically modulated in time domain. 
Unlike the spatial photonic crystal that opens bandgaps in frequency axis, the temporal modulation in PTC causes the interference of time-refracted and time-reflected waves, giving rise to Floquet modes and opening bandgaps in momentum axis \cite{reyes2015observation, lustig2018topological, lyubarov2022amplified, park2022revealing, wang2023metasurface}. 
In particular, the Floquet modes in the momentum bandgaps of PTC support exponential growing (decaying) feature with time \cite{lustig2018topological, lyubarov2022amplified, park2022revealing, wang2023metasurface}, since the temporal modulation breaks time-translation symmetry and hence the system does not conserve energy. 
One interesting question is what would happen in a temporal analogue of the spatial Moir\'e superlattice, i.e., the Moir\'e PTC. 
For example, the spatial Moir\'e superlattice supports flat bands with zero group velocity and leads to spatial localization modes \cite{lou2021theory, dong2021flat, nguyen2022magic}. 
On the contrary, it is excepted that a Moir\'e PTC shall support  extreme narrow bands with giant group velocity and temporal localization modes according to the space-time duality, which yet to be fully studied.

In this paper, we theoretically propose a Moir\'e PTC by extending the concept of Moir\'e superlattice in time domain. 
We study the extreme narrow bands in momentum axis, and explore the temporal localized modes lying at the vicinity the narrow bands, which exhibits periodically self-reconstructing pulses in time. 
The influence of the modulation parameters on the narrowness of the bands and the temporal width of the localized modes are investigated, where we find the potential for developing novel mode-locked lasing mechanism from these temporal localized modes. 
Moreover, we show the temporal width of the generated pulses can be further tuned by cascading Moir\'e PTCs with different configurations. 
Our work therefore provides new insight in wave manipulations by time-varying systems when the concept of Moir\'e superlattice is introduced. 

\begin{figure}[htbp]
    \centering
    \includegraphics[width=8.6cm]{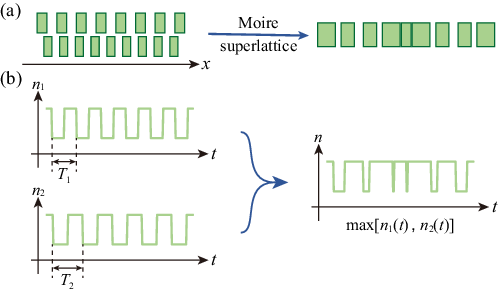}
    \caption{Schematic illustration of the construction of a Moir\'e superlattice in (a) space domain and (b) time domain.}\label{figure.1}
\end{figure}

To construct such a Moir\'e PTC, one can first recall the formation of a spatial one-dimensional (1D) Moir\'e superlattice in Fig. \ref{figure.1}(a), which is obtained by merging two 1D grating with different spatial periods into a single layer. 
The Moir\'e PTC can then be constructed in a similar way by the superposition of two binary PTC $n_1(t)$ and $n_2(t)$ in Fig. \ref{figure.1}(b) with the modulation periods $T_1$ and $T_2$, respectively. 
Here, we set the minimum value of the refractive index being $n_{\mathrm{min}}=1$ for the simplicity, and maximum value being $n_{\mathrm{max}}=n_{\mathrm{min}}+\Delta n$ with $\Delta n$ being the modulation strength and the filling ratio being 0.5. 
Moreover, we assume the modulation periods of the two binary PTC satisfying $T_1/T_2=N_1/N_2$, where $N_1$ and $N_2$ are integers.
The superposed Moir\'e PTC is then governed by spatially-uniform refractive index $n(t)=\mathrm{max}[n_1(t),n_2(t)]$ with modulation period $T=N_1 T_2=N_2 T_1$.

\begin{figure}[htbp]
    \centering
    \includegraphics[width=8.6cm]{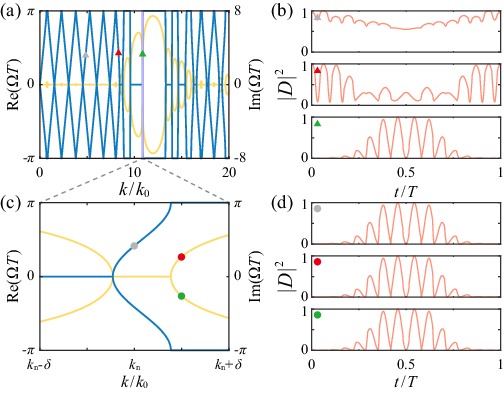}
    \caption{(a) Bandstructure of the Moir\'e PTC with $N_1=15$, $N_2=16$, $n_{\mathrm{max}}=2$, and $n_{\mathrm{min}}=1$. (b) Intensity of the electric displacement field $|D|^2$ for the modes denoted in (a). Zoom in figure of the bandstructure in (a) in the vicinity of the narrow band, where $\delta=0.0048k_0$. (d) Intensity of the electric displacement field $|D|^2$ for the modes denoted in (c).}\label{figure.2}
\end{figure}

We then show a typical bandstructure of such Moir\'e PTCs, which can be obtained from the transfer matrix over one modulation period $T$ with Floquet theorem being applied (see Supplemental Material for detailed derivations \cite{supple}). 
The bandstructure of a Moir\'e PTC with $N_1=15$, $N_2=16$, $n_{\mathrm{max}}=2$, and $n_{\mathrm{min}}=1$ is plotted in Fig. \ref{figure.2}(a), where the blue (yellow) lines denote the real (imaginary) part of eigenvalues $\mathrm{Re}(\Omega)$ [$\mathrm{Im}(\Omega)$]. 
One can clearly see an extreme narrow band represented by the violet line occurs in the vicinity of $k_{\mathrm{n}}=10.8672k_0$ with a superluminal group velocity, where $k_0=2\pi/Tc$ and $c$ is the speed of light in vacuum. 
Note that there are other narrow bands located at both sides of this extreme narrow with larger band widths. 
We clarify that the superluminal group velocity does not violate Einstein's causality. 
Although the group velocity can be superluminal, the information velocity will never be faster than the speed of light as demonstrated in Ref. \cite{pan2023superluminal}. 

We then compared the modes of the Moir\'e PTC with momentum $k$ residing away from the narrow band (gray triangle), close to the narrow band (red triangle) and on the narrow band (green triangle) in Fig. \ref{figure.2}(b). 
For the mode located away from the narrow band, the intensity of the electric displacement field $|D|^2$ is almost evenly distributed over the entire modulation period $T$, where $D$ satisfies $D(t+T)=D(T)$. 
In comparison, the mode located close to the narrow band tends to localized at $t=0, T$. 
The localization behavior becomes stronger for mode located in the narrow band, exhibiting a periodically
self-reconstructing pulses in time, which indicates such temporal localization behavior is a unique feature of the narrow band in Moir\'e PTC. 
We further consider the modes in the vicinity of the narrow band as shown in Fig. \ref{figure.2}(c) with the corresponding $|D|^2$ being displayed in Fig. \ref{figure.2}(d). 
One can see that the modes located at the momentum gap (red and green circles) also exhibits the temporal localization behavior as the one in the narrow band (gray circle), except they have imaginary eigenvalues, indicating an exponential growing (decaying) feature during evolution.

\begin{figure}[bp]
    \centering
    \includegraphics[width=12.9cm]{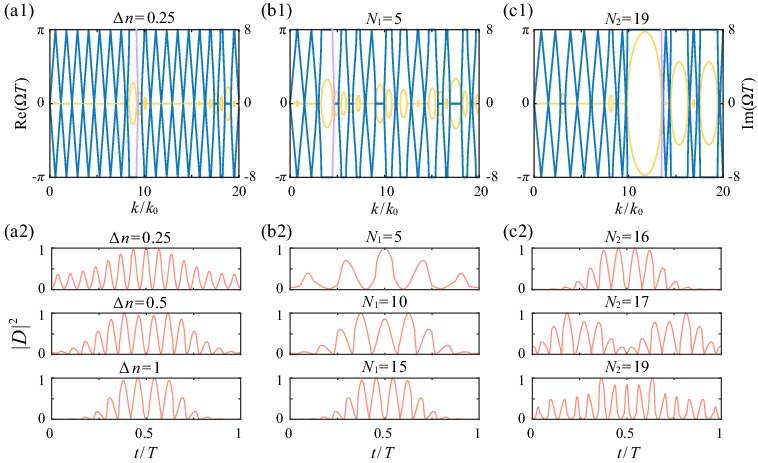}
    \caption{(a1) Bandstructure of the Moir\'e PTC with $N_1/N_2=15/16$ and $\Delta n=0.25$. (a2) Typical temporal profile of the modes at the narrow band [violet line in (a1)] with $N_1/N_2=15/16$ and various $\Delta n$. (b1) Bandstructure of the Moir\'e PTC with $N_1/N_2=5/6$ and $\Delta n=1$. (b2) Typical temporal profile of the modes at the narrow band [violet line in (b1)] with $\Delta n=1$, $N_1+1=N_2$ and $N_1$ varying. (c1) Bandstructure of the Moir\'e PTC with $N_1/N_2=15/19$ and $\Delta n=1$. (c2) Typical temporal profile of the modes at the narrow band [violet line in (c1)] with $\Delta n=1$, $N_1=15$ and various $N_2$.}\label{figure.3}
\end{figure}

Next, we investigate how the modulation parameters influence the narrowness of the band and the growing (decaying) feature temporal localized modes. 
We first consider the effect of the modulation strength $\Delta n$. 
In Fig. \ref{figure.3}(a1)-(a2), we fix $N_1=15$, $N_2=16$, $n_{\mathrm{min}}=1$ and vary $\Delta n$. 
Compared with the bandstructure in Fig. \ref{figure.2}(a), one sees the imaginary eignvalues $\mathrm{Im}(\Omega)$ becomes smaller and the width of the narrow band near $k=10k_0$ is wider with $\Delta n=0.25$ in Fig.\ref{figure.3}(a), which indicates a weaker localization of the temporal localized mode [see the top panel of Fig. \ref{figure.3}(b)]. 
A larger modulation strength $\Delta n$ can open wider momentum gaps \cite{lustig2018topological} which compresses the narrow bands, leading to a stronger localization of the temporal localized modes as shown in Fig. \ref{figure.3}(b). 
We then consider the influence of the modulation pattern of the Moir\'e PTC. 
In Figs. \ref{figure.3}(b1)-(b2), we fix $N_1+1=N_2$ and tune only $N_1$. 
As $N_1$ increases, which results in a more complex modualtion pattern, the localization of the temporal localized modes becomes stronger and the bandwidth of narrow band is smaller [see Fig. \ref{figure.2}(a) and Fig. \ref{figure.3}(b1)], which indicates a narrower band leading a stronger localization in time domain. 
However, this relation is not always true. 
As we show in Fig. \ref{figure.3}(c1)-(c2), where we fix $N_1=15$ and change $N_2$. 
One finds that although increasing $N_2$ gives a more complex modulation pattern and a narrower band [see Fig. \ref{figure.3}(c1) for $N_2=19$], the temporal profile of the modes in Fig. \ref{figure.3}(c2) are different from the ones we study above. 
In other words, we find that such Moir\'e PTCs also has a magic configuration inheriting from the 1D spatial Moir\'e superlattice \cite{nguyen2022magic, wang2020moire, hong2022flatband} where the narrow band gets to be extreme flat.

\begin{figure}[tbp]
    \centering
    \includegraphics[width=8.6cm ]{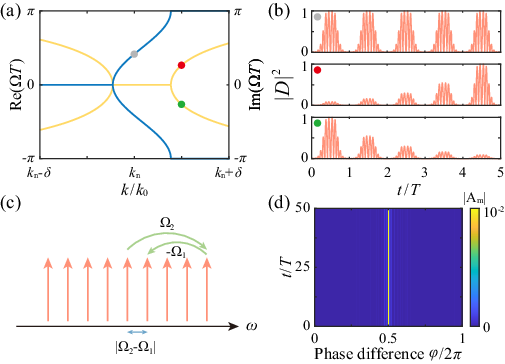}
    \caption{Bandstructure of the Moir\'e PTC with parameters same as Fig. \ref{figure.2}(c). (b) Temporal variation of $|D|^2$ for the modes denoted by the gray, red and green circles in (a), respectively. (c) Illustration of the mode-locking mechanism. (d) Saturated distribution of $|A_m|$ versus the phase difference $\varphi$ between adjacent frequency components and the evolution time $t$ with the maximum value being 1.}\label{figure.4}
\end{figure}

We further explore the dynamics of the temporal localized modes found in Moir\'e PTCs. 
In Fig. \ref{figure.4}(b) we show the temporal evolution of the modes denoted in Fig. \ref{figure.4}(a) within 5 modulation periods from numerical simulations. 
One finds that the modes located at the narrow band (grey circle) and the momentum gap (red and green circles) both shows a periodically temporal localization behavior. 
This is particularly interesting for the modes located at the momentum gap compared with the ones in conventional binary PTCs \cite{lustig2018topological, lyubarov2022amplified, sadhukhan2023defect, sadhukhan2024bandgap}.
For the modes located at the momentum gap in a binary PTC, the growing (or decaying) modes always dominated during the evolution. 
On the contrary, the intensity of the modes in a Moir\'e PTC monotonically increases until it reaches its peak and then decreases within a modulation period $T$, while they still show a globally exponential growing (or decaying) trend throughout the evolution. 
Moreover, we investigate a mode-locking mechanism in such a Moir\'e PTC as a temporal analogue of the one in spatial Moir\'e superlattice \cite{ma2023twisted} as illustrated in Fig. \ref{figure.4}(c). 
For a binary modulation $n_1(t)$ [$n_2(t)$], the Floquet modes are delocalized in time domain, and only the Floquet modes with frequency spacing $\Omega_1$ ($\Omega_2$) can be coupled. 
After the second binary modulation being introduced, the Floquet modes spaced by $|\Omega_2-\Omega_1|$ in frequency can also couple with each other, which leads to the mode-locking in frequency space and the temporal localization. 
We further verify such a mode-locking mechanism by the numerical simulation within 50 modulation periods. 
Here, the amplitude of the frequency components $A_m$ are examined, which are extracted from $D(t)$ in each modulation period by Fourier transform. 
Here, $m$ is an integer denoting the index of the frequency component. 
In Fig. \ref{figure.4}(d), we plot the saturated distribution of $|A_m|$ versus the phase difference $\varphi=\textrm{Arg}(A_m)-\textrm{Arg}(A_{m+1})$ between adjacent frequency components  and the evolution time $t$, where the mode denoted by the gray circle in Fig. \ref{figure.4}(a) is excited. 
One sees the distribution of the frequency components $|A_m|$ is invariant at each phase difference during the evolution, indicating a mode-locking signature hidden in Moir\'e PTCs. 
Therefore, the proposed Moir\'e PTCs holds the potential for developing novel mode-loced lasers by utilizing the modes located at the momentum gaps near the narrow band [e.g., red circle in Fig. \ref{figure.4}(a)], which extract energy from the temporal modulation instead of a gain medium \cite{lustig2018topological,lyubarov2022amplified}.

\begin{figure}[tbp]
    \centering
    \includegraphics[width=12.9cm]{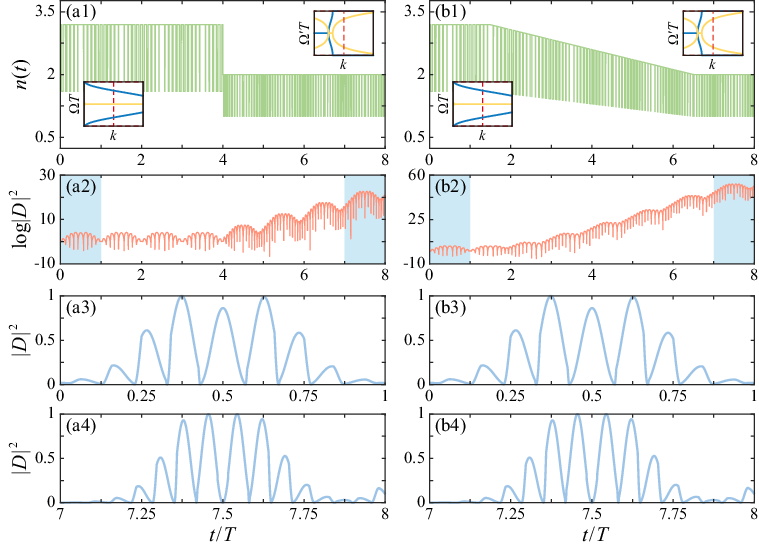}
    \caption{(a1)-(b1) Modulation pattern of the Moir\'e PTC for $t\in[0,8T]$. Insert figures show the corresponding bandstructures for $t\in[0,T]$ and $t\in[7T,8T]$, respectively. (a2)-(b2) Temporal variation of $\mathrm{log}|D|^2$ for the modulation patterns in (a1) and (b1), respectively. Temporal variation of $|D|^2$ for (a3)-(b3) $t\in[0,T]$ and (a4)-(b4) $t\in[7T,8T]$, respectively.}\label{figure.5}
\end{figure}

Moreover, as the modulation pattern influence the temporal profile of the modes, one can dynamically change the temporal width of the generated pulses by varying modulation pattern. 
In Figs. \ref{figure.5}, we show two ways to change the time width of the temporal localized mode by varying the modulation pattern. 
In Fig. \ref{figure.5}(a) for the first way, we set $n_{\mathrm{max}}=3.19$, $n_{\mathrm{min}}=1.595$, $N_1/N_2=10/11$ for $t\in[0,4T]$, and $n_{\mathrm{max}}=2$, $n_{\mathrm{min}}=1$, $N_1/N_2=15/16$ for $t\in[4T,8T]$, where the corresponding bandstructures $\Omega(k)$ and $\Omega'(k)$ are shown in insert figures. 
For another way illustrated in Fig. \ref{figure.5}(b), we set the same parameters as Fig. \ref{figure.5}(a) for $t\in[0,1.5T]$ and $t\in[6.5T,8T]$, while varying $n(t)$, $N_1$ and $N_2$ linearly for $t\in [1.5T,6.5T]$. 
The temporal variation of $\mathrm{log}|D|^2$ are shown in Figs. \ref{figure.5}(a2) and (b2), respectively, where the momentum of the excited modes are denoted by the red dashed line in Figs. \ref{figure.5}(a1) and (b1). 
One can see that the time width of the temporal localized modes in \ref{figure.5}(a4) and (b4) are shortened compared with results in  Figs. \ref{figure.5}(a3) and (b3). 
Although there is difference in the amplification ratio, one finds the temporal profile of $|D|^2$ for these two ways are identical as shown in Figs. \ref{figure.5}(a4) and (b4) for these two ways. 
Such time evolution can be understood by the projection of states. 
As the system holds spatial translational symmetry and breaks time-translation symmetry, where the momentum of states is conserved, the states on $\Omega(k)$ shall project to the states on $\Omega'(k)$ with the same momentum $k$. 
Therefore, as long as the bandstructures of the last modulation periods in two ways are the same, the final temporal profile of the modes does not change.

In summary, we theoretically explore Moir\'e PTCs which support extreme narrow bands. 
We show such narrow bands can support temporal localized modes and explore how the modulation parameters modify the narrowness of the band and the temporal localization. 
The self-reconstructing temporal localized modes indicates the mode-locking mechanism in frequency space, which points to the potential application of  mode-locked lasers with tunable time width of the emitted pulse. 
It might be challenging to realize such Moir\'e PTCs in optical regime as it requires ultrafast temporal modulation \cite{zhou2020broadband, pacheco2022time, lustig2023time, saha2023photonic}. 
Nevertheless, the proposed constructions of Moir\'e PTCs are not limited in photonic system, but can be extended to other physical systems including water waves \cite{bacot2016time, apffel2022experimental}, electric circuits \cite{wang2023metasurface, moussa2023observation, galiffi2023broadband}, ultracold atoms \cite{dong2024quantum}, acoustic waves \cite{kim2024temporal}, and time-multiplexed pluses \cite{ye2023reconfigurable, yu2024dirac}. 
Our work points new direction in wave manipulations with time-varying systems, which may further extend to spatiotemporal  media \cite{caloz2019spacetime, caloz2019spacetime1, engheta2020metamaterials, galiffi2022photonics, yin2022floquet, li2022single}, leading to localized modes in both space and time domain. 
Moreover, the superluminal propagation of modes on the narrow bands \cite{zou2024momentum} may lead to intriguing opportunities in highly efficient optical communication and computation.

\begin{acknowledgments}
The research was supported by National Key Research and Development Program of China (No. 2023YFA1407200 and No. 2021YFA1400900), National Natural Science Foundation of China (12122407 and 12192252).
\end{acknowledgments}

\bibliography{ref}
\end{document}